\documentclass{PoS}
\newcommand{\preprintline}{\newline
\vskip -4.2cm
\rightline{\parbox{4cm}{\large\rm DESY 05-181\\ ITEP-LAT/2005-20\\ KANAZAWA/05-12}}
\vspace{3.2cm}}

\PoS{PoS(LAT2005)157}
\FullConference{XXIIIrd International Symposium on Lattice Field Theory\\
         25-30 July 2005\\
         Trinity College, Dublin, Ireland}

\usepackage{graphicx}
\newenvironment{Eqnarray}%
          {\arraycolsep 0.14em\begin{eqnarray}}{\end{eqnarray}}

\newcommand{\bc}{\begin{center}}
\newcommand{\ec}{\end{center}}
\newcommand{\eq}{\begin{equation}}
\newcommand{\ee}{\end{equation}}
\newcommand{\ea}{\begin{Eqnarray}}
\newcommand{\eea}{\end{Eqnarray}}

\title{Critical temperature in QCD with two flavors of dynamical
quarks\thanks{This work is supported by the SR8000 Supercomputer Project of High
Energy Accelerator Research Organization (KEK). A part of
numerical measurements has been done using NEC SX-5 at Research
Center for Nuclear Physics (RCNP) of Osaka University. The
numerical simulations of this work were done using RSCC computer
clusters in RIKEN. We wish to acknowledge the support of the
computer center at RIKEN.
}\preprintline}

\ShortTitle{Critical temperature in QCD with two flavors of dynamical quarks}
\author{V.G.~Bornyakov$^{a,}$\thanks{VGB, MNC and MIP are supported by grants 04-02-16079,
        RFBR-DFG-03-02-04016, DFG-RFBR 436 RUS 113/739/0, CRDF award RPI-2364-MO-02 and MK-4019.2004.2.},
        M.N.~Chernodub$^b$,
        Y.~Mori$^{c\,d}$,
        S.M.~Morozov$^b$,
        \speaker{Y.~Nakamura}$^{\,\,c\,d}$,
        M.I.~Polikarpov$^b$,
        G.~Schierholz$^e$,
        A.A.~Slavnov$^{f,}$\thanks{A.A.S. is supported by grant Scientific School grant 2052-2003.1.},
        H.~St\"uben$^g$,
        T.~Suzuki$^{c\,d,}$\thanks{T.S. is supported by
        JSPS Grant-in-Aid for Scientific Research on Priority Areas 13135210 and (B) 15340073.
}\\
        \llap{$^a$}Institute for High Energy Physics, RU-142284 Protvino, Russia\\
        \llap{$^b$}ITEP, B.Cheremushkinskaya 25, RU-117259 Moscow, Russia\\
        \llap{$^c$}Institute for Theoretical Physics, Kanazawa University, Kanazawa 920-1192, Japan\\
        \llap{$^d$}RIKEN, Radiation Laboratory, Wako 351-0158, Japan\\
        \llap{$^e$}NIC/DESY Zeuthen, Platanenallee 6, D-15738 Zeuthen, Germany\\
        \llap{$^f$}Steklov Mathematical Institute, Vavilova 42, RU-117333 Moscow, Russia\\
        \llap{$^g$}Konrad-Zuse-Zentrum f\"ur Informationstechnik Berlin, D-14195 Berlin, Germany\\
         E-mail:
                \email{yoshi@hep.s.kanazawa-u.ac.jp}}     

\abstract{We present results obtained in QCD with two flavors of
non-perturbatively improved Wilson fermions at finite temperature
on $16^3 \times 8$ and $24^3 \times 10$ lattices. We determine the
transition temperature in the range of quark masses
$0.6<m_\pi/m_\rho<0.8$ at lattice spacing a$\approx$0.1 fm and
extrapolate the transition temperature to the continuum and to the
chiral limits.
We also discuss the order of phase transition.}

\begin{document}


\section{Introduction}
%
Knowledge of the QCD phase structure is important in order to understand the physics of
the early Universe, structure of neutron stars, processes in heavy ion collisions, {\it etc}.
In order to obtain predictions for the real world from lattice QCD, we have to
extrapolate the lattice data to the continuum and to the chiral limit.
Determination of the transition temperature $T_c$ was studied by many groups.
$T_c$ from staggered fermions with 2 and 2+1 flavors of dynamical quarks~\cite{karsch01,fodor04,hyp,milcthermo},
and from Wilson fermions with 2 flavors~\cite{cppacsnf2} was found to be consistent in the chiral limit.
Edwards and Heller~\cite{Edwards:1999mm} determined $T_c$ for $L_t=4$ and 6 using
nonperturbatively improved Wilson fermions. We compute $T_c$ on finer
lattices with $L_t=8$ and 10
and extrapolate our high statistics data to the continuum limit.
Our results for $L_t=8$ at $\beta=5.2$ were
previously
reported in Ref.~\cite{dik05}.

\section{Simulation}

\begin{figure}[!thb]
\bc \includegraphics[angle=0,scale=0.41,clip=true]{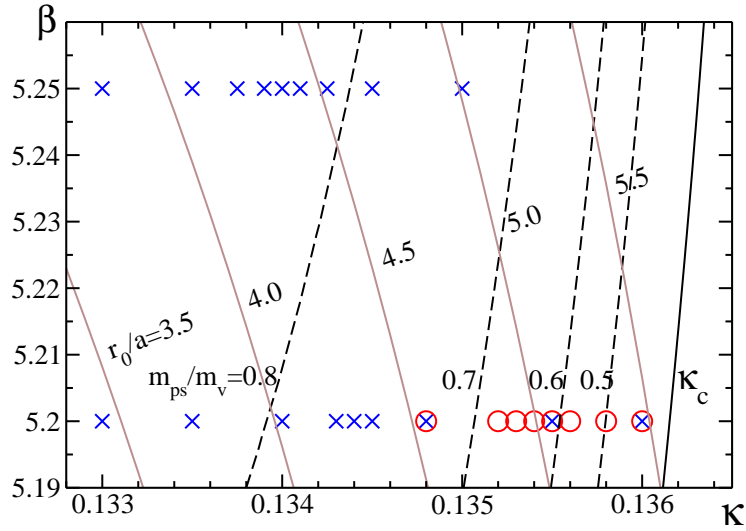} \ec
\caption{Lines of constant $r_0/a$ (solid lines) and constant
$m_\pi/m_\rho$ (dotted lines) at $T=0$. Crosses and circles indicate points
where simulations are done.}
\label{const}
\end{figure}

We use the Wilson gauge action and the non-perturbatively improved Wilson fermions with $c_{sw}$ given in Ref.~\cite{Jansen:1998mx}.
Configurations are generated on $16^3 \times 8$ ($\beta=5.2$ and $5.25$) and
$24^3 \times 10$ ($\beta=5.2$) lattices at various $\kappa$.
The values of $\kappa$ and the corresponding number of configurations for
$24^3 \times 10$ and $16^3 \times 8$ lattices can be found in Table 1.
We use T=0 results obtained by QCDSF-UKQCD to fix the scale.
The contour plot of lines of constant $r_0/a$ and $m_\pi/m_\rho$ is shown in Fig.~\ref{const}.

\vspace{0.8cm}

\bc
\begin{tabular}{|l|r||l|r||l|r|} \hline
\multicolumn{2}{|c||}{$24^3\times 10$}& \multicolumn{4}{|c|}{$16^3\times 8$}                                           \\  \hline
\multicolumn{2}{|c||}{$\beta={{5.2}}$}& \multicolumn{2}{|c||}{$\beta={{5.2}}$}& \multicolumn{2}{|c|}{$\beta={{5.25}}$} \\  \hline
$\kappa$ & \# of conf.                & $\kappa$ & \# of traj.                & $\kappa $    & \# of traj.             \\  \hline
$0.1348$ & $   678 $                  & $0.1330$ & $ 7,120 $                  & $0.1330 $    & $  1,540  $             \\  \hline
$0.1352$ & $  2058 $                  & $0.1335$ & $ 4,500 $                  & $0.1335 $    & $  7,439  $             \\  \hline
$0.1353$ & $  2725 $                  & $0.1340$ & $ 3,000 $                  & $0.13375$    & $  9,225  $             \\  \hline
$0.1354$ & $  3934 $                  & $0.1343$ & $ 6,615 $                  & $0.1339 $    & $ 12,470  $             \\  \hline
$0.1355$ & $  3816 $                  & $0.1344$ & $11,275 $                  & $0.1340 $    & $ 19,479  $             \\  \hline
$0.1356$ & $  1730 $                  & $0.1345$ & $ 9,077 $                  & $0.1341 $    & $ 13,750  $             \\  \hline
$0.1358$ & $  1210 $                  & $0.1348$ & $ 5,813 $                  & $0.13425$    & $  5,155  $             \\  \hline
$0.1360$ & $   617 $                  & $0.1355$ & $ 5,650 $                  & $0.1345 $    & $  2,650  $             \\  \hline
         &                            & $0.1360$ & $ 3,699 $                  & $0.1350 $    & $  1,780  $             \\  \hline
\end{tabular}\\
\vspace{0.2cm}
Table 1: Simulation statistics.
\ec

\bc
\ec

\section{Critical temperature}

\begin{figure}[!thb]
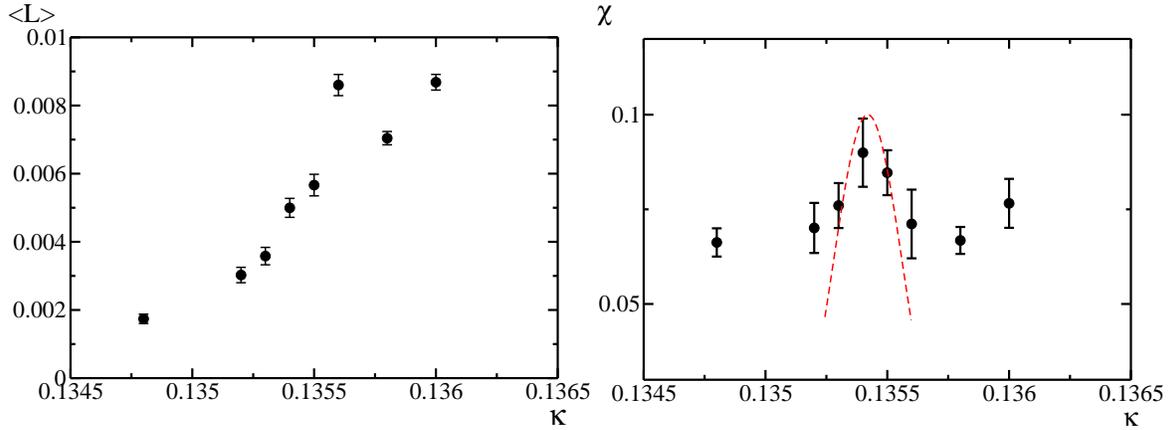

\includegraphics[angle=0,scale=0.3,clip=true]{./figs/polyakov.loop.nonabelian_2410b52.eps}
\includegraphics[angle=0,scale=0.3,clip=true]{./figs/susceptibility.nonabelian_2410b52.eps}
\caption{Polyakov loop (left) and its susceptibility (right) at $\beta=5.2$ on $24^3 \times 10$
lattice.}
\label{NA24_52}
\end{figure}

\begin{figure}[!thb]
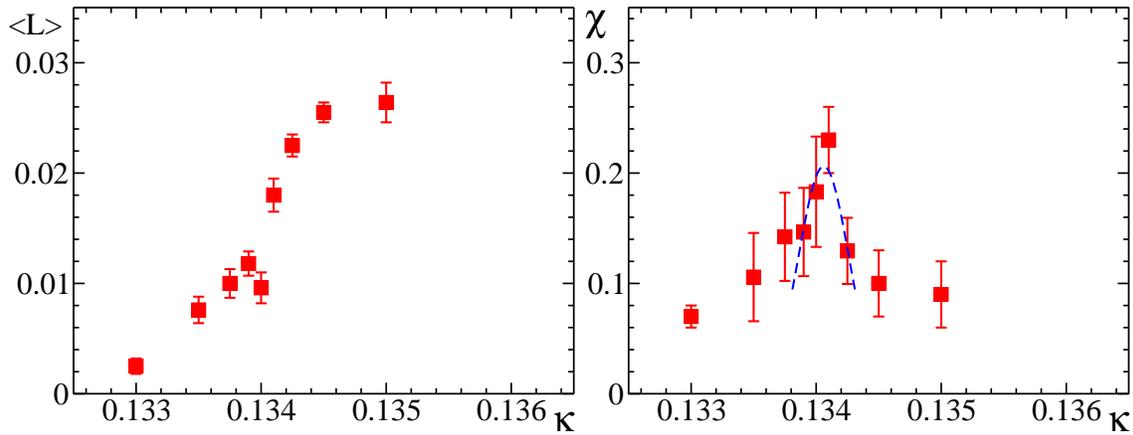

\includegraphics[angle=0,scale=0.34,clip=true]{./figs/polyakov.loop.nonabelian_1608b525.eps}
\includegraphics[angle=0,scale=0.34,clip=true]{./figs/susceptibility.nonabelian_1608b525.eps}
\caption{Polyakov loop (left) and its susceptibility (right) at $\beta=5.25$ on $16^3 \times 8$
lattice.}
\label{NA16_525}
\end{figure}

We use the Polyakov loop susceptibility to determine the transition point.
We identify the critical value of $\kappa$, $\kappa_t$, as a point, where the Polyakov loop susceptibility reaches its maximum.
Applying a Gaussian fit in the vicinity of the maximum (as shown in Figs.~\ref{NA24_52}, \ref{NA16_52} and \ref{NA16_525}) we find
$\kappa_t=0.13542(6)$ at $\beta=5.2$ on $24^3\times 10$,
$\kappa_t=0.13444(6)$ at $\beta=5.2$ on $16^3\times 8$, and
$\kappa_t=0.13406(6)$ at $\beta=5.25$ on $16^3\times 8$.
Using
the Sommer scale
$r_0$=0.5~fm and interpolating $r_0/a$ to the critical point, we obtain for
the critical temperatures
$T_c = 197(2)$MeV (at ${m_\pi /m_\rho} = 0.63(2)$),
$T_c = 212(2)$MeV (at ${m_\pi /m_\rho} = 0.76(1)$), and
$T_c = 217(2)$MeV (at ${m_\pi /m_\rho} = 0.81(1)$), respectively.

\begin{figure}[!thb]
\includegraphics[angle=0,scale=0.34,clip=true]{./figs/polyakov.loop.nonabelian_1608b52.eps}
\includegraphics[angle=0,scale=0.34,clip=true]{./figs/susceptibility.nonabelian_1608b52.eps}
\caption{Polyakov loop (left) and its susceptibility (right) at $\beta=5.2$ on $16^3 \times 8$
lattice.}
\label{NA16_52}
\end{figure}

\begin{figure}[!thb]
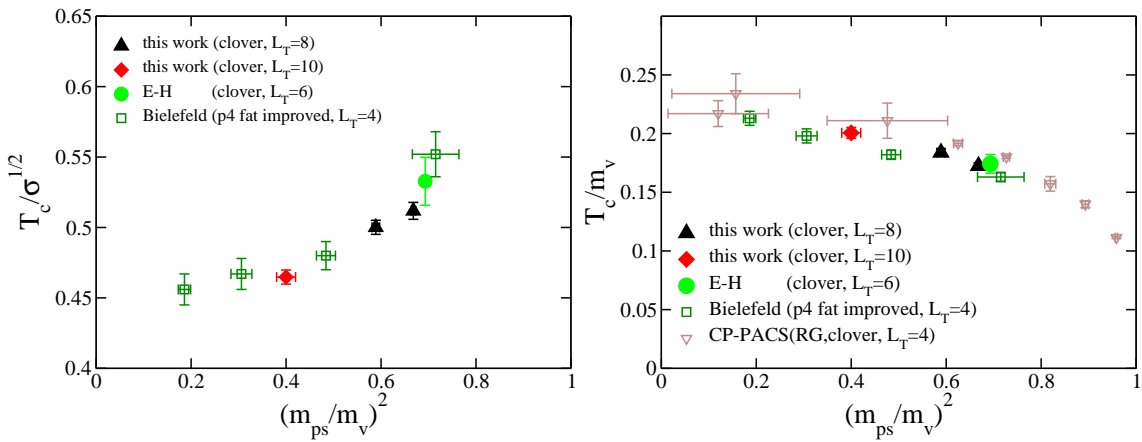

\includegraphics[angle=0,scale=0.31,clip=true]{./figs/Tc.eps}
\includegraphics[angle=0,scale=0.31,clip=true]{./figs/TcMv.eps}
\caption{
(left) The transition temperature $T_c$ in units of the string tension $\sigma(T=0)$ as a function of $m_{\mathrm{ps}}/m_{\mathrm{v}}$
and (right) $T_c$ in units of $m_{\mathrm{v}}$ as a function of $(m_{\mathrm{ps}}/m_{\mathrm{v}})^2$ (right).
The data is from this work (the up triangles and the diamond),
Ref.~\cite{karsch01} (the squares), Ref.~\cite{Edwards:1999mm} (the circle), and
Ref.~\cite{cppacsnf2} (the down triangles).}
\label{Tc}
\end{figure}

In Fig.~\ref{Tc} we compare our result for the critical temperature with those
of Refs.~\cite{karsch01}, \cite{cppacsnf2} and~\cite{Edwards:1999mm},
where we have assumed $\sqrt{\sigma}=425$~MeV, and ${\sf m_{v}}$ and ${\sf m_{ps}}$
are vector meson mass and pseudoscalar mass obtained on the zero-temperature lattice.
Our result is in quantitative agreement with the results of the Bielefeld group and CP-PACS.
This is reassuring, as \cite{karsch01}, \cite{cppacsnf2} and  \cite{Edwards:1999mm}
work at larger lattice spacing.

\bc
\ec

\section{Continuum and chiral limits}


In order to
estimate $T_c$ in the continuum and the chiral limits we apply the extrapolation
formula, which takes into account the lowest order corrections
coming from the finite
lattice spacing and quark masses.
We use two extrapolation functions: the first one is of the bare quark mass type (TYPE 1) and the second one is
of the pseudoscalar mass type (TYPE 2).
%
\ea
&&
 {\bf TYPE~1:~~~~~~}
 T_c r_0 = T_c^{m_q , a \to 0} r_0 +C_a ({a \over r_0})^2
 +C_q ({1 \over \kappa} - {1 \over \kappa_c})^{{1 \over \beta\delta}} \,,
 \label{eq:extr1}\\
&&
 {\bf TYPE~2:~~~~~~}
 T_c r_0 =(T_c r_0)^{m_q , a \to 0} +C_a ({a \over r_0})^2
 +C_q (m_{ps})^{\frac{2}{\beta\delta}} \,.
 \label{eq:extr2}
\eea
where $T_c^{m_q , a \to 0}$ corresponds to the
extrapolated value of the critical temperature, and $\beta$ and $\delta$ are
the
critical indices.
Since $O(a)$ improved Wilson fermions are employed in this studies, we expect corrections of $O(a^2)$ only.

We make an attempt to fit four values for $T_c r_0$ (see Table~2), obtained
at rather large quark masses,
in order to estimate the extrapolation parameters used in Eqs.~(\ref{eq:extr1},\ref{eq:extr2}).
If the transition in two-flavor QCD is
of the
second order,
then
the transition is expected to belong
to the universality class of the $3D$ $O(4)$ spin model with $1/\beta\delta$$\approx$0.54.
Current lattice calculations suggest
that transition in QCD for physical quark masses is not a
true phase transition but a crossover
\cite{karsch01,fodor04,hyp,milcthermo,cppacsnf2,frithjoflat03}.
Since results obtained by CP-PACS indicated
a
good $O(4)$ scaling~\cite{cppacsnf2},
we extrapolate the value of the critical temperature using the value of 0.54 as $1/\beta\delta$.
We rely on the assumption that
the finite temperature phase transition in two-flavor QCD is second order.
The best fit parameters are presented in Table 3.

\vspace{0.2cm}
\bc
\begin{tabular}{|c|c|c|c|c|c|} \hline
$\sf T_c r_0$&$\sf a/r_0 $&$\sf 1/\kappa -1/\kappa_c$&$\sf m_{ps}$ &$\sf L_t$&$\sf \beta$\\  \hline
0.501(5)     &0.200(2)    & 0.040(4)                 &1.58(6)      &10(DIK)  &  5.2      \\  \hline
0.539(5)     &0.232(3)    & 0.094(4)                 &2.25(4)      & 8(DIK)  &  5.2      \\  \hline
0.551(5)     &0.227(2)    & 0.123(4)                 &2.61(4)      & 8(DIK)  &  5.25     \\  \hline
0.57(2)      &0.290(9)    & 0.174(12)                &2.74(5)      & 6(E-H)  &  5.2      \\  \hline
\end{tabular}\\
\vspace{0.2cm}
Table 2:
The available data for the critical temperature $T_c r_0$.
\ec
%
\vspace{0.2cm}
\bc
\begin{tabular}{|c|c|c|c|c|c|} \hline
${\sf 1/\beta\delta}$&$\sf (T_c r_0)^{m_q,a\to 0}$&$\sf C_a$&$\sf C_q$&$\sf \chi^2/dof$&TYPE\\ \hline
 0.54                & 0.439(7)                   &0.1(2 )  & 0.33(3) &  0.09          &  1 \\ \hline
 0.54                & 0.420(7)                   &0.5(1)   & 0.036(2)&  0.04          &  2 \\ \hline
\end{tabular}\\
\vspace{0.2cm}
Table 3:
The best fit parameters for the case if the transition is of second order.
\ec

We get the critical temperature in the continuum and in the chiral limits,
\bc
TYPE1 fitting:  \hspace{2mm} $T_c = 173(3)$MeV, \hspace{2cm} TYPE2 fitting: \hspace{2mm}  $T_c = 166(3)$MeV.
\ec

The value of $T_c$ extrapolated by using
the
TYPE 1 fitting agrees with
the values obtained by other groups,
but the value from
the TYPE 2 fitting is slightly
smaller.
Both fits should give the same result. The difference gives an indication of the systematic error.


Finally, we mention the results of Ref.~\cite{excluO4} which support the first order of the transition.
If the transition is indeed of the first order, then $1/\beta\delta$=1.
In order to test the sensitivity of $T_c$ to the order of the transition,
we also performed a fit with $1/\beta\delta$=1.

\vspace{0.5cm}
\bc
\begin{tabular}{|c|c|c|c|c|c|} \hline
${\sf 1/\beta\delta}$&$\sf (T_c r_0)^{m_q,a\to 0}$&$\sf C_a$&$\sf C_q$&$\sf \chi^2/dof$&TYPE\\ \hline
 1                & 0.486(17)                   &0.0(4 )  & 0.5(1) &  0.11          &  1 \\ \hline
 1                & 0.460(11)                   &0.5(2)   & 0.09(2)&  0.08          &  2 \\ \hline
\end{tabular}\\
\vspace{0.2cm}
Table 4:
The best fit parameters for the case if the transition is of first order.
\ec

\bc
TYPE1 fitting: \hspace{2mm} $T_c = 192(7)$MeV, \hspace{2cm} TYPE2 fitting: \hspace{2mm} $T_c = 182(4)$MeV.\\
\ec

These values are rather large compared to the value obtained from $N_f = 2+1 $ staggered fermions in Ref.~\cite{milcthermo}.
The results of our fits do not discriminate between the first and the second order phase transitions
because of rather large errors in the $T_c r_0$ values.
We are continuing simulations on $24^3 \times 10$ lattice
in order to get a better precision of the $T_c$ value for this lattice.

%
%


\end{document}